\documentclass[aps,prd,a4paper,superscriptaddress,nofootinbib,showpacs,showkeys,amsfonts,amssymb,amsmath]{revtex4}
\usepackage{amssymb,latexsym}
\usepackage{amsmath, amsthm}
\usepackage{amscd}
\usepackage{times}
\usepackage{epsfig}
\usepackage{psfrag}
\usepackage{graphicx}

\newcommand{\be}{\begin{equation}}
\newcommand{\ee}{\end{equation}}
\newcommand{\bea}{\begin{eqnarray}}
\newcommand{\eea}{\end{eqnarray}}

\begin{document}
\title[]{Genericity aspects of black hole formation in the collapse of spherically symmetric slightly inhomogenous perfect fluids}

\author{Seema Satin}
\email{sesatin@mail.tku.edu.tw}
\affiliation{Department of Mathematics, University of Pune, Ganeshkhind, 411007 Pune, India}
\affiliation{Department of Physics, Tamkang University (TKU), New Taipei City, 25137 Taiwan (ROC)}

\author{Daniele Malafarina}
\email{daniele.malafarina@nu.edu.kz}
\affiliation{Center for Field Theory and Particle Physics \& Department of Physics,
Fudan University, 220 Handan Road, 200433 Shanghai, China}
\affiliation{Department of Physics,
Nazarbayev University, 53 Kabanbay Batyr avenue, 010000 Astana, Kazakhstan}

\author{Pankaj S. Joshi}
\email{psj@tifr.res.in}
\affiliation{Tata Institute of Fundamental Research, Homi Bhabha Road, Colaba, Mumbai 400005, India}

\begin{abstract}
We study the complete gravitational collapse of a class of spherically symmetric inhomogeneous perfect fluid models
obtained by introducing small radial perturbations in an otherwise homogeneous matter cloud.
Our aim here is to study the genericity and
stability of the formation of black holes and locally naked singularities
in collapse. While the occurrence of naked singularities is known for many models of collapse, the key issue now in focus is genericity
and stability of these outcomes. Towards this purpose, we study how the introduction of a somewhat general class of small
inhomogeneities in homogeneous collapse leading to a black hole
can change the final outcome to a naked singularity.
The key feature that we assume for the perturbation profile is that of a mass profile that is separable in radial and temporal coordinates. The known models of dust and homogeneous perfect fluid collapse can be obtained from this choice of the mass profile as special cases. This choice is very general and physically well motivated and we show that this class of collapse models leads to the formation of a naked singularity as the final state.

\end{abstract}


\maketitle

\section{Introduction}

One of the issues that is central to black hole physics today is the realistic description of the dynamical formation of black holes. The basic paradigm for black hole formation from gravitational collapse is based on the Oppenheimer-Snyder-Datt (OSD)
\cite{OSD}
model, which describes the complete collapse of a spherical cloud of non interacting (i.e. dust) particles with an homogeneous density profile. In the OSD model the central shell-focusing singularity that forms at the end of collapse is always hidden behind a trapping horizon, thus showing that a Schwarzschild black hole is indeed the final state of collapse. Nevertheless the OSD model is too simplistic, as it assumes pressureless matter and homogeneity. Therefore it is legitimate to wonder how general is the result obtained for homogeneous dust.
This question was at the heart of the original formulation of the Cosmic Censorship Conjecture
\cite{Penrose},
which stated that every natural collapse process must lead to the formation of a singularity that is covered by a horizon at all times, and has been a very fruitful area of research for many years.

Over the past decades there has been substantial analysis of the properties of spherically symmetric dust collapse, both in the homogeneous (OSD) case as well as in the inhomogeneous case (the Lema\`{i}tre-Tolman-Bondi (LTB) collapse model)
\cite{LTB}.
These studies show that the behaviour of the free functions describing the mass and velocity distributions of the particles in the cloud are the crucial elements that determine the final outcome of collapse (see for example
\cite{dust}).
Several studies aimed at clarifying the role of pressures in collapse models have also been done
\cite{pressure},
with some models considering analytical solutions for collapse of perfect fluids with adiabatic equations of state
\cite{CMP},
in order to determine the end states, resulting in a covered singularity or otherwise
(see \cite{review} and references therein).
The key result that emerges from the many models studied over the years is that there are essentially two possible scenarios to describe relativistic gravitational collapse:
\begin{enumerate}
  \item In the black hole case the horizon forms at an outer radius before the formation of the singularity, thus leaving it hidden from far away observers.
  \item In the naked singularity case the singularity and the horizon form at the center at the same time, thus leaving the possibility for geodesics to escape the singularity. In this case the singularity can be either locally or globally naked depending on the behaviour of the horizon close to the boundary of the cloud.
\end{enumerate}
The possibility of real photons escaping the high density region, where the collapsing matter becomes optically thick, that develops close to the formation of the singularity is uncertain. The free mean path of the photon becomes extremely small and, if no other effects intervene, such singularity may not have significant observable effects on the outer regions
 (see for example \cite{Kong-Bambi-Malafarina}).
Therefore the naked singularity in these models must be understood as a region where the fluid model and the relativistic description break down. In this sense they just underline the limits of the classical theory and offer a possible window onto new, possibly explosive, astrophysical phenomena.

The reason why homogeneous models are most often considered relies in the extreme
simplification of Einstein's equations in this case.
When inhomogeneities are introduced the system of Einstein's equations
becomes much more complicated. In order to solve the equations with the inclusion of inhomogeneities and more realistic equations of state
one is naturally led to resort to computer simulations
(see for example \cite{early-numerical} for some early results). Nevertheless one has to remember if the classical signature of a naked singularity is not distinguishable form that of a black hole then there may be no way of gaining much information from numerical simulations (see for example \cite{numerical-ns} for naked singularity formation in numerical simulations or \cite{numerical} for numerical models of collapse that do not present the occurrence of naked singularity).

As of today is still not clear how different kind of pressures affect the final outcome of collapse.
Also how realistic and generic are the known solutions that present naked singularities is still a matter of debate.
One very fruitful approach is based on the study of small perturbations in the initial data of known spherical collapse models.
For example, starting with the homogeneous OSD  model and incorporating
small inhomogeneities in the initial density one can see that the space of final outcomes is equally divided between black holes and naked singularities
\cite{small-pert}.
It is seen that the introduction of small pressure perturbations to dust models
can drastically change the outcome of the collapse from covered to at least locally naked
singularity or vice versa, and that both the black hole and the naked singularity final outcomes are in some sense stable in these cases, with the exception of some special cases.
From this perspective, once suitable definitions of genericity and stability are provided in this context, one can see that naked singularities must be considered a stable and generic outcome of spherically symmetric collapse models just as much as black holes
\cite{genericity}.

Our aim here is to study a class of pressure perturbations in the well-known homogeneous collapse that goes to a
black hole. While this is a general enough class of perturbations, even if it is considered somewhat
limited in the space of all possible perturbations, the main purpose it serves here is to show that
there are physically viable classes of perturbations that change final outcome from black hole to locally naked singularity.
We consider here endless gravitational collapse of a massive spherical star, under the
influence of its own gravitational field. The collapse initiates from
regular initial conditions, where there is no trapping of light and no singularity.
As matter source we consider a perfect fluid where small inhomogeneities are introduced in the initial pressure profile and where no barotropic or polytropic equation of state is imposed.
One can think of the equation of state as being variable from a barotropic type relation at the beginning of collapse to some different relation when the field becomes strong.
Still the fluid source is required to satisfy standard energy conditions.
This work generalizes the earlier results described in
\cite{small-pert} and \cite{genericity},
where certain classes of perturbations in homogeneous collapse leading
to a black hole were studied.

The paper is organized as follows: In section \ref{einst} we review the general setting for relativistic collapse in spherical symmetry. In section \ref{inhom} we study the solution for the system of Einstein's equations under certain physically reasonable assumptions for the fluid source.
Finally in section \ref{conc} we briefly discuss the implications of these result for possible astrophysical observations. Throughout the paper we make use of natural units where $c=G=1$ and we absorb the factor $8\pi$ appearing in Einstein's equations into the definition of energy density and pressure.

\section{The Einstein's Equations for spherically collapsing cloud}\label{einst}

The most general spherically symmetric metric describing a collapsing cloud in co-moving coordinates $r$ and $t$ is given by
\be
 ds^2 = -e^{2 \nu(t,r)} dt^2 + e^{2 \psi(t,r)}  dr^2 + R(t,r)^2 d \Omega \; .
\ee
The energy momentum tensor is $T_\mu^\nu={\rm diag}\{\rho,p_r,p_\theta,p_\theta\}$,
where $\rho $ is the energy density and $p_{\theta}$ and $p_r$ are
radial and tangential pressures.
In the case of a perfect fluid source, where the pressure part of the energy momentum tensor is isotropic, we have $p_{\theta} = p_r=p$.
Einstein's equations in this case take the form
\begin{eqnarray}\label{p}
 p  & = & - \frac{\dot{F}}{R^2 \dot{R}}\; , \\ \label{rho}
 \rho &  = & \frac{F'}{R^2 R'} \; , \\
\nu' & = & - \frac{p'}{\rho + p}\; , \\ \label{eq:nu}
 2  \dot{R}' & = & R' \frac{\dot{G}}{G} + \dot{R} \frac{H'}{H}\; ,  \label{eq:G}
\end{eqnarray}
where $H$ and $G$ are given by
$H := e^{-2 \nu} \dot{R}^2 \mbox{, } G := e^{-2 \psi} R'^2$
and $F$ is the Misner-Sharp mass of the system
\cite{misner}
that is given by
\begin{equation}
F  =   R(1-G+H)\; . \label{eq:motion}
\end{equation}


There are thus five equations in the six unknowns $\rho$, $p$, $F$, $\nu$, $\psi$ and $v$, meaning that
one of the functions is free to choose.
However, the system becomes closed after prescription of an equation of state for the matter content that relates density and pressure.

In the following analysis we proceed without assuming any fixed equation of state. This is due to the fact that at present we do not know the properties of matter under strong gravitational fields, even though we know that the classical description of fluids, as given by barotropic or polytropic equations of state, will not remain unchanged towards the formation of the singularity.
Therefore it makes sense to consider matter fields that have a physically reasonable behaviour in the weak field while not being subject to the constraint of a fixed equation of state when approaching the strong field. In the following we shall consider a variable equation of state where the relation between density and pressure is given by
\be\label{eos}
\lambda(r,t):=\frac{p(r,t)}{\rho(r,t)} \; ,
\ee
with the function $\lambda$ uniquely determined from the choice of the free function of the system. Note that the speed of sound in the cloud, as given by $c_s^2:=dp/d\rho$, is variable but well defined at all times before the formation of the singularity.

There are several assumptions that need to me made in order for the system to be physically viable. We shall discuss them briefly here (refer to
\cite{review} for a more detailed treatment).
We shall consider an initial time $t_i=0$ for which the matter functions $\rho$ and $p$ are regular and present no cusps at the center. This implies that the Misner-Sharp mass takes the form $F(r,t)=r^3M(r,t)$, with $M$ a suitably regular function. Furthermore, we shall consider only collapse scenarios where the weak energy conditions, given by $\rho\geq 0$ and $\rho+p\geq 0$, are satisfied throughout the whole evolution.
The system has a scaling degree of freedom, that can be fixed by choosing the area-radius function $R$ at the initial time. We shall set $R(r,t_i)=r$, thus introducing an adimensional scale factor $v(r,t):=R(r,t)/r$ that describes the rate of collapse.
For a homogeneous perfect fluid we have $p \equiv p(t)$ and $ \rho \equiv \rho(t)$.
This implies a Misner Sharp mass of the form $F(r,t) = r^3 M(t)$ (fulfilling all the regularity conditions), thus making $M$ depend only on `$t$' and $ v \equiv v(t)$.
Using the definition of $G$ and $H$ we can rewrite equation \eqref{eq:G} as
\begin{equation}
\dot{G}=2\frac{\nu'}{R'}r\dot{v}G \; ,
\end{equation}
which can be integrated once we define a function $A(r,t)$ such that
\begin{equation}\label{def:A}
\dot{A} := \frac{\nu'}{R'}r\dot{v} \; .
\end{equation}
Then we get
\begin{equation}\label{A}
G=b(r)e^{2A}\; ,
\end{equation}
  where $b$ is a free function coming from the integration and related to the initial velocity of the infalling particles, usually referred to as the velocity profile. Then regularity requires that near the center $b$ has the form $b(r) = 1 + r^2 b_0(r)$.
 In the following we will consider
 the marginally bound case by imposing $ b_0(r) = 0 $.
 Matching with a Schwarzschild or generalized Vaidya exterior can always be performed at the co-moving boundary radius $r_b$, which corresponds to a shrinking area-radius $R_b(t)=R(r_b,t)$, while matching to a Vaidya solution can be done if one considers an evolving boundary $r_b(t)$
\cite{matching}.

\section{Introducing Inhomogeneities}\label{inhom}

We shall now proceed to integrate the system of Einstein's equations for some inhomogeneous perfect fluid models. Given the absence of a fixed barotropic equation of state, we shall consider the mass profile $M$ as the free function of the system.
Then the energy density $\rho$, the pressure $p$ and the function $\lambda$ in equation \eqref{eos} are uniquely determined by the choice of $M$.
We can consider inhomogeneities in the density and pressure profiles by introducing radial inhomogeneities in the mass profile $M(t) \rightarrow
M(t,r)$. This in turn leads to $ v(t) \rightarrow v(r,t) $. Note that, given the monotonic behaviour of $v$ close to the center and close to the formation of the singularity, we can consider a change of coordinates from the co-moving frame $\{r,t\}$ to the area-radius frame given by $\{r,v\}$ and consider $v$ as a new coordinate to label time. This implies that, for any function $X(r,t)$, we can consider $X(r,v)=X(r,t(r,v))$ so that the radial derivatives in the old coordinates become $X'=X,_r+X,_vv'$, where $X,_r$ is the radial derivative in the new coordinates and $v'$ has now to be understood as a function of $r$ and $v$.

Introducing inhomogeneities in the pressure profile then would result in $p$ and $\rho$ having a radial dependence given by
 \bea
p(v,r) &=& p_0(v) + p_1 (v) r +  \frac{1}{2} p_2 (v) r^2 \; , \\
\rho(v,r) & = & \rho_0(v) + \rho_1(v) r +  \frac{1}{2} \rho_2(v) r^2 \; ,
\eea
where the forms of $p_i(v)$ and $\rho_i(v)$ ($i=0,1,2...$) depend upon the specific choice of the mass function $M$.
For simplicity, we shall now choose the Misner-Sharp mass $F$ in such a way that $M$ be separable in $r$ and $v$, so that has it has the following form
\be \label{m-epsilon}
M(r,v) = m(v)[1 + \epsilon(r)]  \; ,
\ee
where $\epsilon(r)$ can be seen as the radial perturbation to the mass profile and will be assumed to be `small' with respect to $m$, in order to study the departure of the system from the well known homogeneous solutions.
This choice is well justified by the fact that known models, such as homogeneous perfect fluid and inhomogeneous dust, are retrieved from it as special cases (see below). Nevertheless, other classes of collapse that do not rely on this assumption can be investigated
(see for example \cite{CMP} and \cite{small-pert})
and they lead to similar results.
We also assume that $M$ given above
be at least $\mathcal{C}^2 $ in $r$ and  at least $\mathcal{C}^1$ in $v$.
Expanding $M$ in powers of $r$, near $r =0$, and choosing the mass function
up to second order in $r$ we must have
\be \label{eq:massp}
 M(r,v) = M_0(v) + M_1(v) r + \frac{1}{2} M_2(v) r^2 \; .
\ee
Also, since we do not want the initial density and pressure to have cusps at the origin, regularity of the initial data requires that $ M_1 $ vanishes at $r=0$. Thus $ M_1 (v) = m(v) \epsilon'(0) = 0 $, which gives $\epsilon'(0) =0 $. We also assume $ \epsilon(0) = 0 $ for consistency, so that at the center the cloud the system behaves like the homogeneous case described by $m(v)$. This is only a further gauge fixing that does not restrict the generality of the model.
Finally, we have to require $ |M_2| <<  M_0 $, from which we
obtain $\epsilon''(0) << 1 $.

Given the continuity required for $M$, we see that the form of $m(v)$ can be taken as
\be
m(v) = m_0 + m_1 v \; .
\ee
From the above definitions, once we expand pressure and density near $r=0$, we obtain
\begin{eqnarray}
 p(v,r)
 &=& - \frac{m(v),_v}{v^2} - \frac{1}{2} \frac{m(v),_v}{ v^2} \epsilon''(0) r^2 \; ,\\
\rho(v,r)
& = & \frac{3 m(v)}{v^3} + \frac{1}{2} \frac{(5 m(v) \epsilon''(0)) }{ v^2} r^2 \; .
\end{eqnarray}
Since we assume, for a realistic matter model, that the density decreases away from the center, it is clear that we must then require $\epsilon''(0) < 0$.
From the above, we can see that the choice of the mass function implies a choice of the density and pressure profiles.
Also, if we want to express the relation between $\rho$ and $p$ via a variable equation of state, we can write
\be
\rho=\frac{3M+rM,_r-vM,_v}{v^2(v+rv')}-p \; ,
\ee
and thus equation \eqref{eos} implies that the proportionality function between $\rho$ and $p$ is given by
\be
\lambda(r,v)=-\left(1+\frac{3M+rM,_r-vM,_v}{M,_v(v+rv')}\right)^{-1} \; .
\ee
With the above choice of $M$, we see that for $m_1\neq 0$ at the initial time $v=1$ we have
\be
\frac{1}{\lambda}=-\frac{3(m_0+m_1)}{m_1}-\frac{m_0+m_1}{m_1}\frac{r\epsilon'}{1+\epsilon} \; ,
\ee
and thus we retrieve a linear equation of state for the unperturbed case $\epsilon=0$ (when $m_1=0$ the model reduces to the pressureless LTB collapse). Also note that the unperturbed equation of state $\lambda_0(v)=\lambda(0,v)$ goes from an initial value of $\lambda_0(1)=-m_1/3(m_0+m_1)$ to a final dust-like behaviour with $\lambda_0(0)=0$ (for $m_1=-m_0/2$ collapse starts with a radiation-like behaviour).
By simplifying equation (\ref{eq:motion}), using the form of $G$ given by equation \eqref{A}, we can obtain the equation of motion for the system as
\be \label{motion}
\dot{v} = - e^\nu \sqrt{\frac{M}{v} + \frac{ be^{2A} -1}{r^2}}\; ,
\ee
which, once solved for a given choice of the free functions $M$ and $b$, allows to solve the system of Einstein's equations completely.
As said before, in the following we assume the marginally bound velocity profile given by $b(r) = 1$.
To evaluate the solution of the above equation we need the explicit expressions for $ \nu $ and $ A $ in terms of the only free function left.
From equations \eqref{eq:nu} and \eqref{def:A} we get
\bea \label{nu2}
 \nu(r,v) &=& \int_0^r \frac{M,_{vr} v + (M,_{vv} v - 2 M,_v) v'}{(3 M + r M,_r - M,_v v) v} R'd\tilde{r} \; , \\ \label{A2}
A(v,r) &=& \int_v^1 \frac{M,_{vr} v + ( M,_{vv} v - 2 M,_v) v'}{(3M + r M,_r -v M,_v ) v }  r dv \; .
\eea
Given the expansion for $M$, we obtain the corresponding expansion for $ A(v,r)$ near $ r=0 $ as
$A(v,r)  = A_0(v) + A_1(v) r + A_2(v) r^2 + A_3(v) r^3 + A_4(v) r^4+...$, from which we see that using the form of the mass profile given by equation \eqref{eq:massp} implies $ A_0 = A_1= A_3 =  0 $ and
\begin{eqnarray}
A_2(v) & = & \int_v^1 \frac{2 M_{2,v}}{ (3 M_0 - M_{0,v}) } dv= \frac{2}{3} \frac{m_1 \epsilon''(0)}{m_0} (1-v) \; .
\end{eqnarray}
Then we can invert equation \eqref{motion} to obtain $t(r,v)$ as
\be
t(r,v) = t_i + \int_v^1 \frac{e^{-\nu} \sqrt{v} }{\sqrt{M+ 2A_2 v+ 2 r^2A_4v }}dv\; .
\ee
Given the regularity of the functions involved, the solution $t(r,v)$ is in general at least $\mathcal{C}^2$ near the singularity and therefore can be expanded as
\be \label{expansion}
t(v,r) = t(0,v) + \chi_1(v) r + \chi_2(v) r^2 + o(r^3) \; ,
\ee
where $\chi_1 := (dt/dr)_{r=0}$ and $\chi_2 :=  1/2(d^2t/dr^2)_{r=0}$.
The singularity curve $t_s(r)$, representing the time at which the shell labelled by $r$ becomes singular, can be written as
\be
t_s(r) = t(r,0) = t_i +  \int_0^1 \frac{e^{-\nu} \sqrt{v} }{\sqrt{M+ 2 A_2 v+ 2 r^2 A_4v}}dv\; ,
\ee
and, according to equation \eqref{expansion}, it can be expanded as
\be
t_s(r) = t_0 + r \chi_1(0) + r^2 \chi_2 (0) + o(r^3)\; .
\ee
To obtain the expressions for $\chi_1$ and $\chi_2$ we must derive $t(r,v)$ with respect to $r$. Then from
\be
\frac{dt}{dr} = \int_v^1 \left[\frac{ \nu' e^{- \nu} \sqrt{v}}{\sqrt{M +2 A_2v +2 r^2 A_4v}}
- \frac{1}{2} \frac{ e^{-\nu}( M,_r  + 4 r A_4v) \sqrt{v}}{(M + 2 A_2v + 2 r^2 A_4v )^{3/2}}\right]dv\; ,
 \ee
evaluated at $r=0$, we obtain the following expression for $\chi_1$:
\be
\chi_1(v) = - \frac{1}{2}\int_v^1 \frac{M_1\sqrt{v}}{(M_0 + 2 A_2v)^{3/2}} dv \; .
\ee
For simplicity we shall assume that the fourth order term of the expansion of $A(v,r)$ be negligible and thus we will take $A_4=0$. Then, similarly we obtain the expression for $\chi_2$ as
\be\label{chi2}
\chi_2 (v) =  \int_v^1\left [\frac{3}{8} \frac{M_1 ^2}{(M_0+ 2 A_2v)^{5/2}} -\frac{\nu''}{\sqrt{M_0 + 2 A_2v}} -\frac{1}{2}\frac{M_2+ 2 A_2^2v}{(M_0 + 2 A_2v)^{3/2}} \right]\sqrt{v} dv \; .
\ee
Since $M_1 = 0$ we finally get $\chi_1(0)=0 $. Therefore the first non vanishing coefficient in the expansion of the singularity curve is the second order term $\chi_2(0)$. Then $\chi_2$ is given by equation \eqref{chi2} evaluated for $v=0$ and gives
\be
\chi_2 (0) = -\frac{1}{2} \int_0^1 \frac{2\nu''(M_0+ 2 A_2v ) +
M_2 + 2A_2^2v}{ (M_0 + 2 A_2v )^{3/2}}\sqrt{v}dv\; .
\ee

As it was shown in \cite{JD},
it is the value of $\chi_2(0)$ that determines the nature of the singularity and its local visibility. Positivity of $\chi_2(0)$ implies that the singularity curve is increasing in the co-moving time $t$ and thus the singularity forms at first at the shell $r=0$. Positivity of $\chi_2(0)$ is also the necessary and sufficient condition for the apparent horizon to be increasing in $t$ and it is possible to show that this is a necessary and sufficient condition also for null geodesics to escape the central singularity that forms at $t_s(0)$.
Now, by using the expression for $M_0$ and $M_2$, in terms of $m$ and $\epsilon$, as given by $M_0 = m(v) = (m_0 + m_1 v)$ and $M_2 = m(v) \epsilon''(0) $ we can obtain the expressions for $\nu''$ and $A_2$,  up to first order in $m_1/m_0$, from equations \eqref{nu2} and \eqref{A2} as
\bea
\nu'' &=& \frac{1}{3} \frac{m_1}{m_0} \epsilon''(0) v  \; ,\\
A_2 &=& \frac{2}{3} \frac{m_1}{m_0} \epsilon''(0) (1-v) \; ,
\eea
from which we get
\be
\chi_2 (0) = \int_0^1  \frac{ -\frac{1}{3} \frac{m_1}{m_0} \epsilon''(0) v^{3/2}[ m_0(1+ \frac{m_1}{m_0} v) + \frac{4}{3} \frac{m_1}{m_0} \epsilon''(0) v
(1-v) ] - \frac{1}{2}[ m_0 (1 + \frac{m_1}{m_0} v ) \epsilon''(0) ] v^{1/2}}{
m_0^{3/2} [ 1 + \frac{m_1}{m_0} v + \frac{4}{3} \frac{m_1}{m_0}
\frac{\epsilon''(0)}{m_0} v (1-v) ]^{3/2} }dv \; .
\ee
Keeping terms up to order $m_1/m_0$ and neglecting higher order in the same, the
expression reads
\be \label{eq:intchi}
\chi_2(0) = - \int_0^1 \frac{\epsilon''(0)}{m_0^{1/2}}\left[ \frac{v^{1/2}}{2}
+ \frac{m_1}{m_0} \left( \frac{7}{12} v^{3/2} -
 \frac{\epsilon''(0)}{m_0} ( v^{3/2} - v^{5/2}) \right) \right] dv \; ,
\ee
which, after performing the integration, gives
\be
\chi_2(0) =  - \frac{\epsilon''(0) }{ m_0^{1/2} } \left( \frac{1}{3} + \frac{m_1}{m_0}
\left( \frac{7}{30}   - \frac{4}{35} \frac{ \epsilon''(0)}{m_0} \right) \right) \; .
\ee
Since we expect $|\epsilon''(0)|/m_0 < 1$ we can ignore the last term in the
above, due to smallness of the multiplying factor, and write
\be \label{eq:chi2}
\chi_2(0) = - \frac{\epsilon''(0)}{ 3 \sqrt{m_0}} \left(1 +  \frac{7}{10}
\frac{ m_1}{m_0} \right) \; .
\ee
As we have discussed earlier, physically reasonable density profiles require that $\epsilon''(0) < 0$, then the sign of $\chi_2$ is decided by the quantity in brackets.
It is clear that for small departures from the homogeneous perfect fluid model we must have $|m_1/m_0| < 1$, regardless of the sign of $m_1$ in the above. It follows that the quantity in brackets is always positive, which in turn implies that $\chi_2(0) $ is always positive. We then conclude that, for scenarios described by small deviations from the homogeneous perfect fluid model as described above, collapse results in the formation of a locally naked singularity.

By looking at the final expression \eqref{eq:chi2} we can see how the above model can be related to the well known homogeneous perfect fluid and dust models.

\begin{enumerate}
\item We can retrieve the OSD collapse model in the case when $m_1$ and $\epsilon''(0)$ (and all the
 higher derivatives of $\epsilon(r)$ at $ r=0$) vanish. In this case the mass function reduces to $M=m_0$, which implies that $\chi_2(0)=0$ and we obtain a simultaneous singularity, resulting in a black hole final state.
\item
In the same manner, we can obtain the homogeneous perfect fluid model by imposing that $\epsilon''
(0) $ and all higher derivatives of the same at $r = 0$ are zero. In this case $M$ becomes a function of $t$ only, through $v(t)$, and again the singularity is simultaneous thus resulting in the formation of a black hole.
\item To obtain the inhomogeneous dust collapse described by the LTB model we must impose $m_1 = 0 $. Then the mass profile becomes a function of $r$ only, given by $M=m_0(1+\epsilon(r))$ and we get $\chi_2 (0)  =  -  \epsilon''(0)/ (3 \sqrt{m_0})$ which, for $ \epsilon''(0)<0$, leads to the formation of a naked singularity.
\end{enumerate}
Thus, from the above framework, we recover the widely studied cases of spherical collapse that lead to the formation of a black hole,
namely homogeneous dust and homogeneous perfect fluid collapse, as well as the inhomogeneous dust model leading to a naked singularity.
The important result of the above analysis is that we immediately see how the addition of a small pressure perturbation to a known
collapse model can change the outcome from black hole to a locally naked singularity.

\section{Conclusion}\label{conc}

In the present paper we have investigated how the introduction of small pressure perturbations in known collapse scenarios of dust and homogeneous perfect fluid affects the final outcome of collapse. We have chosen a very general class of physically valid mass profiles, given by a separable mass function.
Our calculations show that inhomogeneous perfect fluids can collapse to a naked singularity and that the homogeneous case is somehow `special', in the fact that it leads to a simultaneous singularity, a result that agrees with what was previously found in
\cite{small-pert}.

This analysis suggests that towards the final stages of collapse a star could reach a stage with arbitrarily high densities at the center before the formation of the trapping horizon. This may have important astrophysical consequences, as the region where the classical description breaks down might be causally connected with the outside universe, thus changing drastically the classical picture for collapse (see for example \cite{quant}).
In recent years a lot of attention has been devoted to the theoretical study of observational features of naked singularities and exotic compact objects
(see for example \cite{obs}).
At present the question of the nature of massive and supermassive compact objects observed in space remains open
(see for example \cite{kundt}).
Future observations will provide experimental evidence on the final state resulting from the complete gravitational collapse of a star. Only then we will have information on the still open question whether black holes are the only necessary outcome of collapse, or if other possibilities, as allowed by theoretical models such as the ones discussed here, do occur in nature.
Our result supports the idea that gravitational collapse to black holes, as described in the OSD model, might not be the most general paradigm to describe the final moments of the life of a star. Different density and pressure profiles contribute to the occurrence of naked singularities within theoretical models. Such singularities, that signal a breakdown of the classical relativistic description, might in turn indicate the existence of an observable window into the physics that dominates at small scales when the gravitational field is large.




\end{document}